**Title:** Extending Inferences from Randomized Controlled Trials to Target Populations: A Scoping Review of Transportability Methods

**Authors:**

Guanbo Wang, PhD[1], Ting-Wei Ernie Liao, MD[2], David Furfaro, MD[3], Leo Anthony Celi, MD, MPH, MSc[3,4,5,*], Kevin Sheng-Kai Ma, DDS[1,6,7,*]

**Affiliations:**

[1]Department of Epidemiology, Harvard T.H. Chan Schol of Public Health, Boston, MA 02115

[2]Section of Cardiac Electrophysiology, Division of Cardiovascular Medicine, Department of Medicine, University of Pennsylvania School of Medicine, Philadelphia, PA 19104

[3]Division of Pulmonary, Critical Care and Sleep Medicine, Beth Israel Deaconess Medical Center, Boston, MA 02215

[4]Laboratory for Computational Physiology, Massachusetts Institute of Technology, Cambridge, MA 02139

[5]Department of Biostatistics, Harvard T.H. Chan School of Public Health, Boston, MA 02115

[6]Department of Dermatology, Massachusetts General Hospital, Boston, MA 02114

[7]Center for Global Health, Perelman School of Medicine, University of Pennsylvania, Philadelphia, PA 19104

[*]These authors contributed equally as senior authors

**Abstract:**

**Objective:** Randomized controlled trial (RCT) results often inform clinical decision-making, but the highly curated populations of trials and the care provided during the trial are often not reflective of real-world practice. The objective of this scoping review is to identify the ability of methods to transport findings from RCTs to target populations.

**Study design:** A scoping review was conducted on the literature focusing on the transportability of the results from RCTs to observational cohorts. Each study was assessed based on the methodology used for transportability and the extent to which the treatment effect from the RCT was estimated in the target population in observational data.

**Results:** A total of 15 published papers were included. The research topics include cardiovascular diseases, infectious diseases, psychiatry, oncology, orthopedics, anesthesiology, and hematology. These studies show that the findings from RCTs could be translated to real-world settings, with varying degrees of effect size and precision. In some cases, the estimated treatment effect for the target population were statistically significantly different from those in RCTs.

**Conclusion:** Despite variations in the magnitude of effects between RCTs and real-world studies, transportability methods play an important role in effectively bridging the RCTs and real-world care delivery, offering valuable insights for evidence-based medicine.

**Keywords:** transportability, randomized controlled trials, observational studies, generalizability, causal inference.

**Introduction:**

Randomized controlled trials (RCTs) are considered the gold standard in medicine for estimating the effects of treatments.(1) However, the patient populations and the care they receive in RCTs often do not represent the target population and their care in the real world. RCTs are conducted with more care than is provided in real-world settings and subject the results to observer bias. Therefore, the conclusions derived from RCTs may not translate to the target populations.(2) Statistical methods, such as transportability analyses (also known as generalizability, calibration, and population adjustment), can be used to translate effect size estimates of treatments from an RCT population to a target population.(3) Such analyses can inform clinicians about the causal effects of a treatment in a target population.

Different methods have been developed for transportability analyses.(4) In causal inference, three types of methods have been proposed. The first is the weighting method, which models the probability of a subject participating in the RCT and weights the subjects in the RCT using this probability.(5) Another is outcome regression, which fits the regression model from the RCTs to estimate conditional means, then marginalizes over or standardizes to the target population covariate distribution by predicting counterfactuals.(6) A more advanced method involves modeling both the probability of the treatment given the patient covariates and the probability of the outcome. The causal treatment effect estimate is consistent if either the probability is consistently estimated. Such estimators are called doubly robust estimators. Note that all the methods assume the individual patient data (for the covariate, but not necessarily for the outcome and treatment) are available in the target population. In addition, the methods work with the transportability assumption, which requires the counterfactual outcome distribution to be exchangeable in two populations (treated and untreated), conditional on the covariate and treatment. When the models are correctly specified, the outcome regression and doubly robust estimators are often more precise.

Given the rapid development of the methods designed to provide a theoretical framework for transporting findings from RCTs to target populations, the aim of the present study was to survey and describe the different methods as well as their clinical use cases.

**Methods:**

This paper provides a scoping review of the approaches to translate treatment effect estimates from RCTs to other target populations represented in observational data such as registries, claims data, and electronic health records. We retrieved relevant articles in PubMed, IEEExplore, and Embase in January 2024. We defined eligible studies as those that (1) included both data from RCTs and observational data, (2) reported statistical inferences of the findings, and (3) were published in English peer-reviewed journals. The search strategy was developed using the following Boolean logic: ("transport" OR "transportability" OR "generalize" OR "generalizability") AND ("randomized controlled trials" OR "randomized controlled trial" OR "randomized clinical trials" OR "randomized clinical trial" OR "clinical trials" OR "clinical trial") AND ("observational studies" OR "observational study" OR "cohort studies" OR "cohort study" OR "real-world" OR "real world").

For each eligible study, we extracted the following information: (1) the names, inclusion and exclusion criteria, interventions, and outcomes from the RCTs; (2) the data sources, the definition of the target population, interventions, and outcomes from observational studies (target population); (3) the scale of the target parameter, such as hazard ratio (HR), relative risk, odds ratio (OR), mean ratio, risk ratio (RR), risk difference (RD), or mean difference. If the target parameter was a causally interpretable parameter, we determined (1) whether the number of groups was compared in the transportability analysis, (2) whether the heterogeneity of treatment effect was investigated, (3) what statistical methods or models were used, (4) whether the data privacy was an issue when applying the transportability methods, (5) whether machine learning was employed when implementing the methods, (6) what findings were obtained by utilizing the methods, and (6) whether the estimated parameter of the target population was statistically significantly different from the RCT.

**Results:**

***Selected studies***

The selected studies included two studies in cardiovascular diseases that focused on heart failure (HF) and Dual Antiplatelet Therapy (DAPT) effects (5, 7), two in infectious diseases that looked at infections from human immunodeficiency virus (HIV) and the prevention of cytomegalovirus (CMV) infections in liver transplant recipients (8, 9), three studies in psychiatry on HIV/acquired immunodeficiency syndrome (AIDS)-related suicidal thoughts, depression in HIV patients, and recent-onset schizophrenia (10-12), two studies on non-small cell lung cancer (13, 14), one study on hip fracture (15), one study on bone marrow transplantation (16), and one study on addiction medicine that assessed the impact of stimulant use on treatment initiation (17). In addition, three methodological studies that applied their methodological findings to a coronary artery disease registry were included (3, 6, 18) (**Table 1-2**).

***Transporting research findings in cardiovascular diseases from RCTs to target populations***

In a study on HF, regression-based and inverse odds of sampling weights (IOSW) were used to transport findings from the RCT, Carvedilol or Metoprolol European Trial (COMET) to a large heart failure registry. A similar analysis was conducted for another RCT, the Digitalis Investigation Group trial (DIG) (7). In both RCTs, the primary outcome was all-cause mortality, and the composite outcome was all-cause mortality or hospitalization.

The observational data (target population) were from a Scottish HF registry, comprising 8,012 patients with reduced ejection fraction. Compared to the patients included in the RCTs, the patients in the registry were older, with severe renal dysfunctions, and received higher dosages of loop diuretics.

In COMET, 3,029 patients with a left ventricular ejection fraction of 35% or less across 15 European countries were followed for four years. The RCT showed that among the trial population, the ORs of carvedilol versus metoprolol for the primary and composite outcomes are 0.83 (0.74, 0.93) and 0.94 (0.86, 1.02), respectively. The transportability analyses showed that the ORs in the target population are 0.62 (0.39, 0.99) and 0.87 (0.59, 1.30), respectively.

In DIG, 6,800 patients with chronic HF were followed for three years. The RCT showed that among the trial population, the ORs of digoxin versus placebo for the primary and composite outcomes are 0.99 (0.91, 1.07) and 0.75 (0.69, 0.82), respectively. The transportability analyses showed that the ORs in the target population are 1.06 (0.92, 1.21) and 0.73 (0.64, 0.83), respectively.

The above results were obtained using IOSW; regression-based methods showed similar results. The study also performed subgroup analyses and found similar treatment effect estimates in high-risk and low-risk registry patients. The study found that the point estimates of the treatment effects were different but not statistically significant between the populations of RCTs and the registry. In addition, the estimates in RCTs were more precise as regards confidence interval than those in the target population. Using statistical transportation techniques, such as IOSW, the study showed the practicality of transporting the treatment effects from an RCT to a target population.

In a study on extended-DAPT following percutaneous coronary intervention (PCI), inverse odds of trial participation weighting methods were used to transport findings from the DAPT Study to a more contemporary cohort from the National Cardiovascular Data Registry CathPCI Registry. (5) The DAPT Study enrolled 8,864 US patients who received drug-eluting stents (DES), and compared 30 versus 12 months of DAPT. The observational data (target population) included 568,540 patients from the CathPCI Registry between July 2016 and June 2017. Compared to the trial population, registry patients were older, had more comorbidities, and were more likely to receive 2nd-generation DES.

In the DAPT Study, the treatment effect for extended-duration DAPT demonstrated a decrease in stent thrombosis (1.01% (–1.41% to –0.54%), major adverse cardiac and cerebrovascular events (MACCE) (1.90% (–2.95% to –0.92%), and an increase in severe or moderate bleeding (0.89% (0.33% to 1.51%). However, after reweighting to represent the registry cohort, these effects were attenuated: stent thrombosis (–0.40% (–0.99% to 0.15%), MACCE (–0.52% (–2.62% to 1.03%), while the increase in severe or moderate bleeding worsened (1.15% (–0.08% to 2.45%).

While the original DAPT Study suggested benefits of prolonged therapy in reducing ischemic events, when adjusting for the characteristics of a more diverse patient population, these benefits were attenuated, while the risk of bleeding increased. These findings underline the importance of considering patient and procedural characteristics when applying results in RCTs to real-world practice.

In a series of three methodological studies, the comparative effectiveness of coronary artery surgery in conjunction with medical therapy versus medical therapy alone for chronic coronary artery disease was investigated using the Coronary Artery Surgery Study (CASS) database.(3, 6, 18) The CASS database comprised 780 who consented participants from a larger cohort of 2,093 eligible patients. The observational component included the remaining 1,319 patients who declined randomization. 1,686 patients with complete baseline covariate data were included in the transportability study. This subset was divided into 731 randomized individuals (368 undergoing surgery and 363 receiving medical therapy) and 955 nonrandomized patients (430 undergoing surgery and 525 receiving medical therapy). The studies utilized a variety of estimators, including augmented IPSW, IPSW, doubly robust estimators, and outcome model-based standardization results (OM). For the purpose of this review, we selectively presented the OM results as a representative example to illustrate the comparative analyses. Readers seeking an in-depth examination of all results generated by different estimators are encouraged to refer to the original studies.

In the first study, the analysis underscored the impact of treatment modalities on the 10-year mortality risk among randomized participants versus the entire trial-eligible cohort.(6) For randomized individuals, the combined treatment of coronary artery surgery and medical therapy was associated with a reduction in 10-year mortality risk to 17.4% (13.5%, 21.3%), compared to 20.1% (15.9%, 24.3%) for medical therapy alone, resulting in a RD of -2.7% (-8.5%, 3.0%) and a RR of 0.86 (0.59, 1.14). An OM analysis for all trial-eligible individuals estimated a 10-year mortality risk of 18.5% (14.3%, 22.6%) for surgery and 20.0% (15.9%, 24.0%) for medical therapy, leading to an RD of -1.5% (-7.2%, 4.2%) and an RR of 0.93 (0.64, 1.21).

The second study looked at people who took part and people who did not. It found that the 10-year risk of death was 17.4% (13.6% to 21.4%) for people who had surgery and 20.4% (16.3% to 24.6%)) for people who were only on medical therapy (RD of -3.0% (-8.7%, 2.7%); RR of 0.85 (0.62, 1.15). (3) The OM analysis for people who did not take part showed that the 10-year risk of death was 18.9% (13.9%, 22.7%) for surgery and 20.1% (15.9%, 24.5%)) for medical therapy, with an RD of -1.3% (-7.9%, 4.2%)) and an RR of 0.94 (0.65, 1.24).

In the third study, the differential benefits of treatments based on a history of myocardial infarction (MI) were reported.(18) For individuals with a history of MI, the RD from trial data was -4.4% (-12.1%, 3.4%), while the OM estimates suggested an RD of -5.1% (-13.1%, 2.9%). In contrast, for individuals without a history of MI, the trial data suggested an RD of 1.4% (-6.2%, 8.9%), and the OM provided a similar estimate of 1.5% (-6.0%, 9.1%). In a nonrandomized subset of the target population of trial-eligible individuals, the subgroup-specific average treatment effects for previous MI presented similar findings. The OM estimates for the nonrandomized subset were RD of -4.7% (-12.8%, 3.5%) for individuals with a history of MI and 1.4% (-6.4%, 9.2%) for those without. These analyses highlight the impact of previous MI on the effectiveness of the treatment, as estimated in both randomized and nonrandomized cohorts within the eligible population in the study.

Using various transportability estimators, a non-trivial treatment effect difference is suggested by these studies between RCTs and observational studies. It was highlighted that the observed therapeutic benefits of coronary artery surgery in conjunction with medical therapy, as opposed to medical therapy alone for chronic coronary artery disease, were consistent across different settings, although the precision of estimates varied. The employment of augmented IPSW, IPSW, doubly robust estimators, and OM in these analyses demonstrated the feasibility and utility of statistical transportability techniques to bridge the gap between RCT findings and real-world applications, thereby enhancing the external validity of RCTs.

***Transporting research findings in infectious diseases from RCTs to target populations***

In a study on HIV treatment, inverse probability-of-selection weights was employed to test the robustness of findings from the AIDS Clinical Trial Group (ACTG) 320 trial, comparing a highly active antiretroviral therapy (HAART) to a largely ineffective combination therapy.(9) The observational data were derived from the US population infected with HIV in 2006, as estimated by the Centers for Disease Control and Prevention (CDC). The RCT enrolled 1,156 HIV-infected adult men and women across the United States in 1996, randomly assigning 577 to HAART and 579 to the control group and followed participants for 52 weeks. The observational data (target population) was characterized by differences in age, sex, and race compared to the trial participants. Compared to the trial participants, the target population was expected to be more diverse in terms of age distribution, with a larger proportion of younger individuals, and racial composition, with a higher percentage of Black, non-Hispanic individuals.

The intent-to-treat analysis of the ACTG 320 trial found HR=0.51 (0.33, 0.77) of AIDS or death for the treatment group compared to the control group. However, when adjusting for differences in age, sex, and race/ethnicity between the trial sample and the target population using inverse probability-of-selection weights, the resulting HR=0.57 (0.33, 1.00) was not far off from the RCT estimate.

***Transporting research findings in organ transplantation from RCTs to target populations***

In a study focusing on the role of preemptive therapy (PET) for CMV disease prevention in CMV high-risk donor seropositive/recipient seronegative (D+R-) liver transplant recipients, the authors compared data from a RCT with data from real-world clinical practice.(8) The RCT comprised 100 liver transplant recipients who were CMV D+R- and were part of the PET group in the CAPSIL trial. These participants came from various transplant medical centers across the United States. In contrast, the observational study conducted a retrospective analysis at a single U.S. academic medical center, involving 50 adult CMV D+R- liver transplant recipients undergoing their first liver transplant, with PET as standard clinical care. The primary outcome of the study was the occurrence of CMV infections one year post-transplantation. In the real-world cohort, 8% (4 out of 50 patients) developed CMV disease, similar to the 9% (9 out of 100 patients) in the CAPSIL clinical trial group (p=1.0). Additionally, the rate of breakthrough CMV disease during the 100-day PET period was low and comparable in both cohorts, with 4% in the real-world group and 3% in the CAPSIL cohort. This similarity in outcomes between the RCT and real-world settings suggests that PET is an effective and feasible approach for preventing CMV disease in high-risk populations, both in RCTs and in real-world clinical practice.

***Transporting research findings in psychiatry from RCTs to target populations***

In a study on schizophrenia, regression-based methods and inverse probability of sample weighting (IPSW) were employed to transport findings from the Disease Recovery Evaluation and Modification (DREaM) RCT to a real-world population of adult Medicaid beneficiaries with recent-onset schizophrenia.(10) The DREaM trial, involving 132 patients aged 18 to 35 diagnosed with schizophrenia or schizophreniform disorder who had experienced their first psychotic episode within the last 24 months, compared the effects of oral antipsychotics (OAP) and paliperidone palmitate (PP), a long-acting injectable antipsychotic. Patients in the RCT were divided into three treatment groups: OAP-OAP, OAP-PP, and PP-PP. This analysis also included data from a cohort of 1,000 Medicaid patients with schizophrenia, identified from the IBM MarketScan Medicaid Managed Care (MMC) database between 2015 and 2019. Over the course of 18 months, the RCT found that each patient in the MMC group had 0.83 (standard error (SE) 0.14) psychiatric hospitalizations, 0.43 (SE 0.14) in the unweighted OAP-OAP group, and 0.80 (SE 0.37) in the calibrated OAP-OAP group. The results of the calibrated OAP-OAP group closely resembled those of the real-world MMC cohort, indicating satisfactory calibration. The 18-month cumulative psychiatric hospitalizations were significantly lower in the OAP-PP group compared to the MMC cohort, with a mean difference of -0.77 (−1.08, −0.47). Similarly, the PP-PP group also had significantly lower hospitalizations, with a mean difference of -0.83 (−1.15, −0.60). The study demonstrated a more pronounced and statistically significant reduction in psychiatric hospitalizations in the RCT setting compared to the observational study, indicating that the benefits of PP over OAP in reducing hospitalizations observed in the DREaM trial are overestimated when compared to a real-life group of Medicaid recipients recently diagnosed with schizophrenia.

In a study on the use of efavirenz in HIV treatment, inverse odds weights and multiple imputation methods were employed to transport findings from four RCTs to a cohort from the Centers for AIDS Research Network of Integrated Clinical Systems (CNICS), comprising 8,291 adults living with HIV who were receiving clinical care at eight academic medical center sites in the U.S.(11) The RCTs, enrolling 3,949 patients across 68 US sites, found an increased risk of suicidal thoughts or behaviors associated with efavirenz, with HR=2.3 (1.2, 4.4). However, when these findings were transported to the observational CNICS cohort, the estimated HR=1.8 (0.9, 4.4), though the incidence rate difference (IRD) remained similar between the trial participants and the observational study target population (trials: IRD=5.1 (1.6, 8.7); observational study: IRD=5.4 (-0.4, 11.4)). This suggests that while both the RCTs and the observational study observed an association between efavirenz use and an increased occurrence of suicidal thoughts and behaviors, the effect size was more pronounced in the RCT environment.

In a study on the effects of a depression treatment intervention among HIV-infected adults, regression-based methods and IPSW were utilized to transport findings from the SLAM DUNC trial to a target population represented by the Centers for AIDS Research's (CFAR) Network of Integrated Clinical Systems (CNICS) cohort. (12) The SLAM DUNC trial enrolled 304 HIV-infected patients aged 18-64 from four US infectious disease clinics, focusing on individuals with a confirmed diagnosis of major depressive disorder, excluding those with treatment-resistant depression or acute psychiatric episodes, between 2010 and 2014. The participants were randomly assigned to either receive the measurement-based Care (MBC) intervention or enhanced usual care. The observational data (target population) were derived from the CNICS cohort, comprising 3,176 patients who were both HIV-infected and depressed, receiving routine care at various clinical sites in the United States.

In the RCT, the MBC intervention was associated with a 3.6-point improvement in depression on the Hamilton Depression Rating scale after 6 months (-5.9, -1.3). However, when IPSW was applied to transport the effect to the target population, the intervention effect was attenuated, leading to an average improvement of 2.4 points (-6.1, 1.3). The study found the intervention's magnitude of effect to be stronger in the RCT settings compared to the observational study setting.

***Transporting research findings in oncology from RCTs to target populations***

In the study assessing the generalizability of the Chinese ORIENT-11 trial to a US cohort with advanced non-small-cell lung cancer, IPSW was employed to integrate findings from the ORIENT-11 trial, which involved 397 patients, with observational data from a de-identified advanced NSCLC database within the Flatiron Health electronic health record, encompassing 557 patients from the United States.(13) The trial involved the use of sintilimab combined with pemetrexed and platinum (SPP) as an intervention, compared to a placebo mixed with pemetrexed and platinum (PPP). After statistical adjustments to align the trial data with US population features, the study concluded that the superiority of SPP over PPP in terms of progression-free survival (PFS) advantage remained unchanged. Specifically, after IPSW adjustment, the median PFS was 8.1 months for the SPP arm compared to 4.9 months for the PPP arm, resulting in HR=0.42 (0.34, 0.53). The observed result aligns with the original HR=0.48 (0.36, 0.64) from the ORIENT-11 trial before adjustment. This consistency in outcomes before and after adjustment indicates that the effects observed in the Chinese ORIENT-11 trial could be applicable to the US population with similar clinical characteristics. The findings indicated that the treatment efficacy observed in the RCT setting was generalizable to a real-world US setting, demonstrating the practicality of using results in RCTs to inform treatment effects in broader populations.

In a study on advanced stage non-small cell lung cancer (NSCLC), the inverse propensity score reweighting methodology was used to transport findings from the pivotal phase III JMDB trial, which compared cisplatin plus pemetrexed with cisplatin plus gemcitabine, to a targeted real-world population derived from the observational FRAME study.(14) The JMDB trial enrolled 1209 chemotherapy-naïve patients with stage IIIB or IV NSCLC and an Eastern Cooperative Oncology Group (ECOG) performance status of 0 or 1 who were over 18 years of age with histologic or cytologic diagnosis of non-squamous histology. The observational data (target population) were from the FRAME study, which included 948 patients with a similar stage of NSCLC and ECOG performance status with histological confirmation. The primary outcome was median overall survival (OS). The original unweighted analysis of the JMDB trial showed HR=0.85 (0.75, 0.97) of OS for pemetrexed versus gemcitabine. After applying the inverse propensity score reweighting, HR=0.91 (0.62, 1.31), indicating a smaller effect size with greater uncertainty given wider CI. The effective sample size (ESS, n=126) after reweighting was only 10% of the original unweighted sample size (n=1222, pemetrexed (n=608) vs. gemcitabine (n=614)), highlighting the impact of reweighting on the statistical significance of the results given the smaller sample size. This study utilized the inverse propensity score reweighting methodology to address generalizability concerns and estimate the expected treatment benefit in a broader real-world target population. By adjusting for differences in baseline characteristics between trial participants and the real-world target population, the method provided insights into the expected treatment benefits in a broader patient cohort, addressing a common concern among healthcare decision-makers regarding the external validity of RCT evidence.

In a study on the effectiveness of umbilical cord blood (UCB) versus HLA-haploidentical bone marrow (Haplo-BM) transplantation for adults with high-risk hematologic malignancies, the authors included 368 patients from a phase III RCT comparing UCB transplantation with HLA-haploidentical bone marrow (Haplo-BM) transplantation.(16) This RCT included a total of 183 patients who underwent UCB transplants and 154 patients who underwent Haplo-BM transplants. The study was conducted at 33 different centers in the United States from June 2012 to June 2018. Concurrently, observational data from 956 patients, including 195 undergoing two-unit UCB transplants, 358 with Haplo-BM transplants, and 403 with haploidentical peripheral blood (Haplo-PB) transplants, were analyzed. These procedures occurred at 91 centers, 32 of which participated in the RCT. Extended follow-up revealed that the five-year progression-free survival (PFS) rate was 37%, (29%, 45%) for recipients of Haplo-BM transplantation and 29%, (22%, 36%) for those receiving two-unit UCB transplantation (p=0.08). The OS rates were 42%, (33%, 51%) for Haplo-BM and 36%, (29%, 44%) for two-unit UCB transplantation, respectively (p=0.06). Comparing these outcomes with those from observational studies indicated no significant difference in the five-year OS rates between the observational and RCT cohorts for both two-unit UCB (36%, (29%, 44%) vs. 41%, (34%, 48%), p=0.48) and Haplo-BM (42%, (33%, 51%) vs. 47%, (41%, 53%), p=0.80) transplantation groups. This similarity supports the generalizability of the RCT findings to a broader population. However, within the observational cohort, individuals undergoing Haplo-BM transplantation exhibited a significantly higher five-year survival rate than those in the RCT two-unit UCB group (47%, (41%, 53%) vs. 36%, (29%, 44%), p=0.012). Those outside the RCT receiving Haplo-PB transplantation had a higher five-year survival rate of 54% compared to RCT participants and those undergoing Haplo-BM transplantation in the observational studies (HR=0.76, (0.57, 0.99), p=0.044, and HR=0.78, (0.63, 0.97), p=0.027, respectively). Survival outcomes were notably better following Haplo-PB transplantation compared to both the RCT and observational UCB cohorts (HR=0.57, (0.45, 0.74), p<0.0001 and HR=0.63, (0.50, 0.81), p=0.0002, respectively). The analysis indicated that the findings of the RCT were applicable in real-world settings. Nonetheless, Haplo-PB transplantation demonstrated a survival advantage over both UCB and Haplo-BM transplantation in these settings, suggesting a preference for Haplo-PB as a superior option for transplantation in adults with high-risk hematologic malignancies.

***Transporting research findings in orthopedics from RCTs to target populations***

In a study utilizing regression-based and IPSW to transport findings from the RCT on the generalizability of treatment effects observed in a RCT of hip fracture surgery implants to a wider population, data from the World Hip Trauma Evaluation (WHiTE-3) trial were used alongside observational data from two extensive databases: the WHiTE-cohort and the UK National Hip Fracture Database (NHFD).(15) The WHiTE-3 trial, comprising 958 participants, conducted a comparison between modular hemiarthroplasty implants and standard monoblock implants in patients aged 60 years and older who had intracapsular hip fractures in the United Kingdom. The observational data consisted of 2,457 patients from the WHiTE cohort and 190,894 patients from the NHFD. The primary outcomes assessed were health-related quality of life (HRQoL), which was evaluated using the EuroQol (EQ-5D-5L) scale, and the length of hospital stay (LOS). In terms of HRQoL, differences in the estimations of treatment effects between the WHiTE-3 trial and the observational data (including both the WHiTE-cohort and NHFD) were insignificant, with a mere 0.01-point difference in the EuroQol (EQ5D). Regarding LOS, the difference between the WHiTE-3 trial estimate and the WHiTE-cohort was 0.50 days, whereas the difference between the WHiTE-3 trial and the NHFD estimate was -0.47 days. This analysis suggests that the findings from the WHiTE-3 clinical trial were generalizable to the broader UK population with intracapsular hip fractures, demonstrating consistency in the magnitude of effects on the outcomes across the RCT and observational studies.

***Transporting research findings in anesthesiology from RCTs to target populations***

In the study exploring the impact of stimulant use on the initiation of medications for opioid use disorder (MOUD) across different population groups, the authors utilized an integrative data analysis approach, employing IPSW to combine data from two RCTs with observational data from three distinct populations.(17) The RCTs, CTN-0051 X:BOT and CTN-0067 CHOICES, included a total of 673 participants from various inpatient opioid treatment facilities and outpatient HIV clinics across the United States. The observational data encompassed 139 respondents from the National Survey on Drug Use and Health (NSDUH), weighted to represent approximately 661,650 noninstitutionalized, housed adults who identified a need for OUD treatment; 71,751 treatment episodes from the Treatment Episodes Dataset (TEDS), which includes adults entering OUD treatment in facilities receiving public funding across the U.S.; and 1,933 individuals from the Rural Opioid Initiative (ROI), comprising people who misuse opioids and have high rates of injection drug use in rural regions of the U.S.

The results of the RCT showed that the use of stimulants decreased the likelihood of starting MOUD by 32%. (adjusted HR=0.68, (0.49, 0.94). However, this effect showed variability when applied to real-world populations. In the NSDUH cohort, the reduction in MOUD initiation due to stimulant use was 25% (adjusted HR=0.75, (0.48, 1.18), and in the TEDS dataset, the reduction in MOUD initiation due to stimulant use was 28% (adjusted HR=0.72, (0.51, 1.01). In particular, the effect was stronger in the ROI group, where stimulant use led to a 39% drop in the start of MOUD (adjusted HR=0.61, (0.35, 1.06). The study also highlighted a significantly greater negative impact on the initiation of extended-release naltrexone compared to buprenorphine, particularly in rural populations. The study consistently found that stimulant use is negatively associated with the initiation of MOUD across different populations. This negative association was more pronounced in real-world settings, especially in rural communities with limited access to MOUD.

**Discussion:**

We conducted a scoping review of the applications of transportability methods to transport findings from an RCT to a target population represented by observational data. The interplay between epidemiological methodology, health service research, and regulatory frameworks for data acquisition is essential to extend causal inference from RCTs to target populations. This is valuable in understanding treatment effects on populations or scenarios that are impractical for inclusion in RCTs but may be more frequently encountered in real-world clinical care.

Due to the inclusion and exclusion criteria of RCTs, the population in an RCT usually does not represent the general population or a population of interest. Therefore, the conclusions drawn from RCTs may not be directly applicable to general populations or a particular target population. Leveraging the advantages of RCTs in estimating treatment effects and the fact that observational data covers a broader and more relevant population, transportability methods can extend the findings from an RCT to a target population by either re-weighting the population in RCT according to the target population or marginalize the findings into the target population. It was observed that, in most studies included in this scoping review, the estimated target parameters in target populations were different from those presented in RCTs (aligned in the same direction but exhibited a lower magnitude and larger variation), which verified the necessity and utility of the transportability methods to support clinical or policy decision-making. In addition, this work also served as a review of the current landscape of transportability methods and their applications. Overall, choosing transportability techniques requires additional consideration of the differences between the two populations, which would allow for the selection of tailored solutions, ultimately enabling the successful extension of causal inferences from RCTs to the target population.

Limitations of this scoping review include the exclusion of studies that did not incorporate data from either RCTs or real-world sources, such as the study that transports findings from one observational data to another. Non-peer-reviewed works presented in conference abstracts and preprints were likewise not included in this scoping review. Lastly, as more studies are conducted with advanced transportability techniques, the evolving nature of the topic will require future follow-up reviews to assess the applicability of transportability methods in bridging the gap between the world according to RCTs and real-world clinical practice.

In conclusion, many studies have employed transportability methods to extend causal inferences from RCTs to real-world populations. These methods can be implemented to validate the results from the RCTs and/or obtain conclusions about the target population using the data from RCTs and data from the target population without treatment and outcome information.

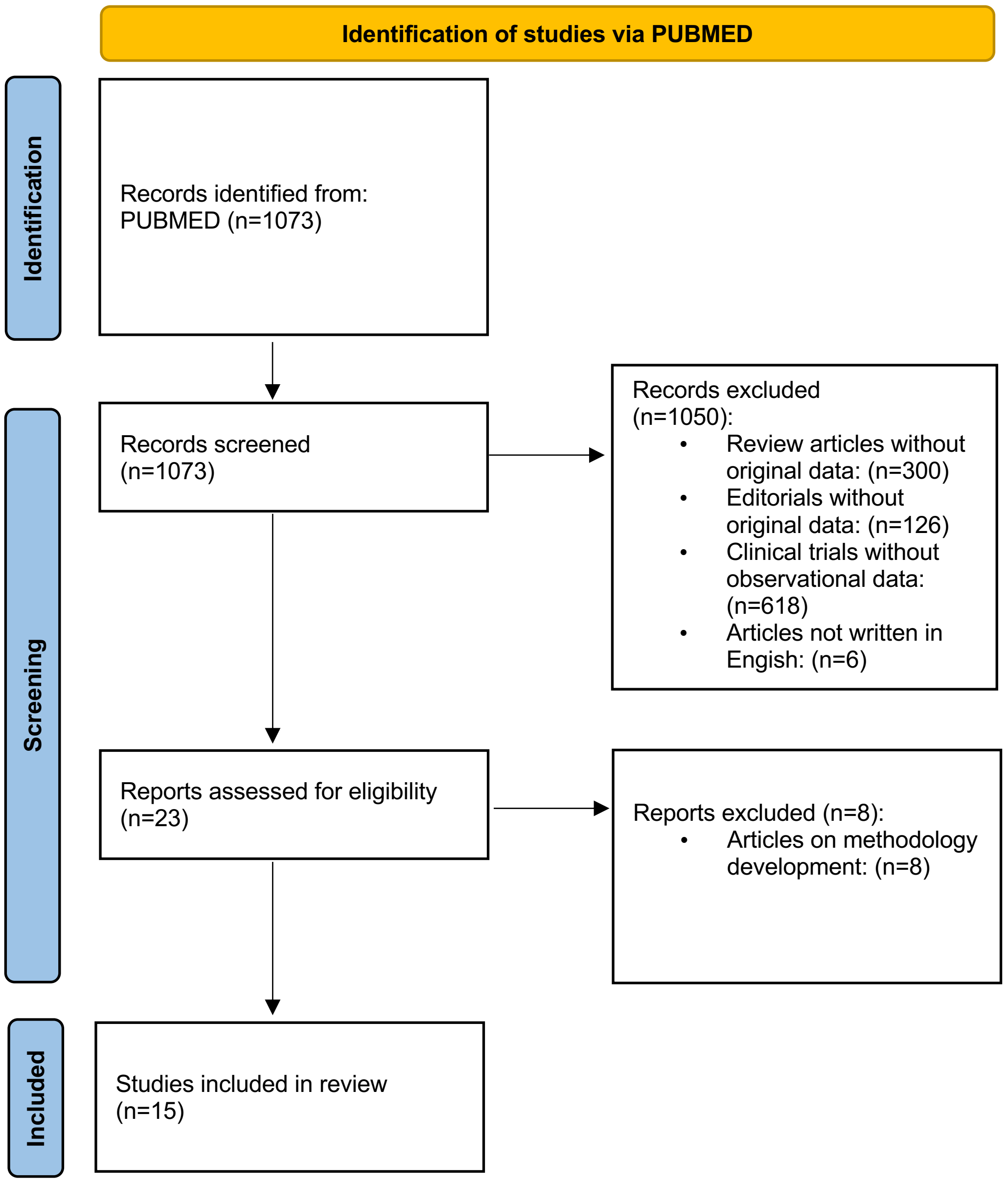

Figure 1. Selection of studies included in the review

Table 1. Randomized clinical trials and target populations in the included studies

| | | Characteristics of the randomized clinical trials | | | | | | Characteristics of the target population | | | | | |
|---|---|---|---|---|---|---|---|---|---|---|---|---|---|
| Study | Country | Number of trials | Name of trial or data source | Definition of population | Intervention or exposure | Control or non-exposed population | Outcomes | Number of sources | Data source | Definition of population | Intervention or exposure | Control or non-exposed population | Outcomes |
| Mollan et al., 2021(11) | USA | 4 | ACTG | - Trial Participants: Enrolled in 2001–2007, ART-naive, at least 18 years of age.<br>- Observational Cohort: US adults who, between 1999 and 2015, began antiretroviral treatment while receiving HIV clinical care at medical centers. | Efavirenz-containing regimens for HIV treatment | Efavirenz-free regimen as first-line ART | Suicidal thoughts/behaviors | 2 | - Trials: ACTG<br>- Observational Cohort: CNICS | Adults with HIV, ART-naive, initiating ART between 1999 and 2015 | Efavirenz-containing regimen | Efavirenz-free regimen | Suicidal thoughts/behaviors |
| O'Donnell et al., 2022(16) | USA | 1 | BMT CTN 1101; NCT01597778 | Adults with high-risk hematologic malignancies | Non-myeloablative transplantation with either two UCB units or HLA-Haplo-BM | Not explicitly stated as a control group, but the study compared outcomes between different intervention groups | 5-year PFS and OS | 3 | The main data sources were 2 clinical trials (BMT CTN 1101; NCT01597778) and the CIBMTR for non-trial data. | Patients aged 18 to 70 years with high-risk acute leukemia or lymphoma, meeting the eligibility of RCT criteria but not enrolled in the trial | Two-unit UCB, Haplo-BM, or Haplo-PB transplantation following the RCT regimen | Each group compared amongst themselves and to each other | 5-year PFS and OS post-transplant |
| Nagasaka et al., 2022(13) | USA | 1 | ORIENT-11 | Patients with untreated locally advanced/metastatic nonsquamous NSCLC | SPP | PPP | PFS, OS, tumor response to therapy, safety outcomes | 1 | ORIENT-11 trial data and a US real-world deidentified advanced NSCLC database | Adults with untreated locally advanced/metastatic nonsquamous NSCLC, based on ORIENT-11 eligibility criteria | SPP | PPP | PFS, OS, tumor response to therapy, safety outcomes |
| Basu et al., 2023(10) | USA | 1 | DREaM | Adult Medicaid beneficiaries with recent-onset schizophrenia | Paliperidone palmitate (PP), a long-acting injectable antipsychotic | OAP therapy | Reductions in psychiatric hospitalizations | 2 | DREaM study data and IBM Truven databases (2015 to 2019) | Patients with schizophrenia enrolled in MMC and diagnosed with schizophrenia using IBM Truven databases | PP for the intervention groups | OAP therapy for the control group | Psychiatric hospitalizations over 18 and 30 months |
| Wei et al., 2023(7) | UK | 2 | COMET, (DIG) | COMET: participants with a LVEF of 35% or less<br>DIG: people with chronic HF | Carvedilol (COMET), Digoxin (DIG) | Metoprolol (COMET), Placebo (DIG) | All-cause mortality; composite outcomes of all-cause mortality or hospitalization (COMET) and | 3 | COMET, DIG trials, and Scottish HF registry | Scottish HF registry patients | Carvedilol (COMET), Digoxin (DIG) | Metoprolol (COMET), Placebo (DIG) | All-cause mortality; composite outcomes of all-cause mortality or hospitalization (COMET) and |
| | | | | | | | HF-related death or hospitalization (DIG) | | | | | | HF-related death or hospitalization (DIG) |
| Lee et al., 2021(15) | UK | 1 | WHiTE-3 trial | Patients aged 60 years and over with intracapsular hip fractures considered suitable for a hemiarthroplasty. | Modern cemented modular polished-taper stemmed hemiarthroplasty (modular arm). | Traditional cemented monoblock (monoblock arm). | HRQoL using the EQ-5D-5L at 4 months postfracture and LOS in days. | 3 | WHiTE-3 trial, WHiTE-cohort, UK NHFD. | Patients aged 60 years and over with intracapsular hip fractures considered suitable for a hemiarthroplasty. | Modern cemented modular polished-taper stemmed hemiarthroplasty (modular arm). | Traditional cemented monoblock (monoblock arm). | HRQoL using the EQ-5D-5L at 4 months postfracture and LOS in days. |
| Cook et al., 2023(17) | USA | 2 | ACTG 320, CTN-0051 X:BOT, CTN-0067 CHOICES | Different populations across trials and datasets: HIV-infected adults (ACTG 320); individuals with OUD in clinical trials and real-world settings (CTN-0051 X:BOT, CTN-0067 CHOICES, NSDUH, TEDS, ROI) | Highly active antiretroviral therapy for ACTG 320; Buprenorphine or extended-release naltrexone for OUD in CTN trials | Largely ineffective combination therapy for ACTG 320; varying by trial and dataset for OUD | Incident AIDS or death within 1 year for ACTG 320; initiation of MOUD for CTN trials and associated analyses | 5 | ACTG 320 trial data, CDC estimates, NSDUH, TEDS, ROI | Participants in two multisite CTN treatment trials: 0051 (X:BOT) and 0067 (CHOICES) | Time-varying stimulant use (cocaine, methamphetamine, or amphetamines) | Participants without stimulant use | Initiation of MOUD |
| Bengtson et al., 2016(12) | USA | 1 | SLAM DUNC trial | HIV-infected depressed adults | MBC intervention | Enhanced usual care for depression | Improvement in depression measured by the HAM-D | 2 | SLAM DUNC trial, the CFAR and CNICS cohort | HIV-infected, depressed adults in routine care in the U.S. | MBC intervention | Enhanced usual care for depression | Improvement in depression measured by the HAM-D |
| Doss et al., 2023(8) | USA | 1 | CAPSIL | CMV D+R- liver transplant recipients | PET for CMV | Participants randomized to PET in the CAPSIL study | CMV disease, biopsy-confirmed acute allograft rejection, retransplant, invasive fungal infections, and death within 1 year post-transplant | 1 | Retrospective assessment at a single US academic medical center | CMV D+R- adults who received a first liver transplant | PET with CMV plasma quantitative PCR for 100 days post-transplant | CAPSIL study PET cohort | Incidence of CMV disease and secondary outcomes (e.g., acute allograft rejection, death) within 1 year post-transplant |
| Dahabreh et al., 2019(6) | USA | 1 | CASS | Patients with chronic coronary artery disease who were eligible for the CASS trial but did not consent to randomization | Coronary artery bypass grafting surgery plus medical therapy | Medical therapy alone | 10-year mortality risk | 1 | CASS data | Individuals with chronic coronary artery disease eligible for CASS, including both randomized and nonrandomized individuals | Coronary artery bypass grafting surgery plus medical therapy | Medical therapy alone | 10-year mortality risk |
| Dahabreh et al., 2020(3) | USA | 1 | CASS | Patients with chronic coronary artery disease who were eligible for the CASS trial but did not consent to randomization | Coronary artery bypass grafting surgery plus medical therapy | Medical therapy alone | 10-year mortality risk | 1 | CASS data | Trial-eligible patients with chronic coronary artery disease not consenting to randomization | Coronary artery surgery plus medical therapy | Medical therapy alone | 10-year mortality risk |
| Robertson et al., 2024(18) | USA | 1 | CASS | Patients with chronic coronary artery disease who were eligible for the CASS trial but did not consent to randomization | Coronary artery bypass grafting surgery plus medical therapy | Medical therapy alone | 10-year risk of death (cumulative incidence proportion) and risk difference in subgroups defined by history of myocardial infarction | 1 | CASS data | Patients with chronic coronary artery disease in subgroups defined by history of myocardial infarction | Coronary artery surgery plus medical therapy | Medical therapy alone | 10-year risk of death and risk difference |
| Cole et al., 2010(9) | USA | 1 | ACTG 320 | HIV-infected adults in the US in 1996 | Highly active antiretroviral therapy | Largely ineffective combination therapy | Incident AIDS or death within 1 year | 1 | ACTG 320 trial data and CDC estimates for HIV-infected individuals in the US in 2006 | Adults infected with HIV in 2006, estimated by the CDC | Highly active antiretroviral therapy | Largely ineffective combination therapy | Incident AIDS or death within 1 year |
| Happich et al., 2020(14) | European Countries | 1 | JMDB trial | Chemotherapy-naïve patients aged ≥ 18 years with stage IIIB or IV NSCLC | Pemetrexed + cisplatin | Gemcitabine + cisplatin | OS | 2 | JMDB trial data and FRAME study | Patients with advanced stage NSCLC | Pemetrexed + cisplatin | Gemcitabine + cisplatin | OS |
| Butala et al., 2022(5) | USA | 1 | DAPT | Patients who received PCI with drug-eluting stents | Extended-duration DAPT for 30 months | Standard-duration DAPT for 12 months | Stent thrombosis, MACCE, MI, bleeding events | 1 | DAPT Study and NCDR CathPCI Registry | | Contemporary population of US patients undergoing PCI with drug-eluting stents from July 2016 to June 2017 | Standard-duration DAPT for 12 months | Stent thrombosis, MACCE, MI, bleeding events |

* ACTG, AIDS Clinical Trials Group; ART, antiretroviral therapy; BM, bone marrow; BMT CTN, Blood and Marrow Transplant Clinical Trials Network; CASS, Coronary Artery Surgery Study; CFAR, Centers for AIDS Research; CIBMTR, Center for International Blood and Marrow Transplant Research; CNICS, Centers for AIDS Research Network of Integrated Clinical Systems; COMET, Carvedilol or Metoprolol European Trial; CTN, Clinical Trials Network; DAPT, Dual Antiplatelet Therapy Study; DIG, Digitalis Investigation Group Trial; DREaM, Disease Recovery Evaluation and Modification; EQ-5D-5L, EuroQol five dimensions questionnaire; HAM-D, Hamilton Depression Rating Scale; HF, heart failure; HRQoL, health-related quality of life; Haplo-BM, HLA-haploidentical bone marrow; LOS, length of hospital stay; MACCE, major adverse cardiac and cerebrovascular events; MI, myocardial infarction; MBC, Measurement-Based Care; MOUD, medication for opioid use disorder; NCDR CathPCI Registry, National Cardiovascular Data Registry CathPCI Registry; NSCLC, non-small-cell lung cancer; NSDUH, National Survey on Drug Use and Health; OAP, oral antipsychotic therapy; OUD, opioid use disorder; PFS, progression-free survival; PCI, percutaneous coronary intervention; PET, preemptive therapy; PP, Paliperidone palmitate; PPP, Placebo + Pemetrexed + Platinum; ROI, Return on Investment; SPP, Sintilimab + Pemetrexed + Platinum; TEDS, Treatment Episode Data Set; UCB, umbilical cord blood.

Table 2. Transportability methods and summary of the included studies

| Study | The scale of the target parameter | Target parameter was a causally interpretable parameter | Number of interventions (including the control) compared in the transportability analysis | Treatment effect heterogeneity was investigated. | Statistical methods or models used | Data privacy was an issue when applying the transportability methods | Machine learning methods were employed | Findings obtained by utilizing the methods | Summary of findings after applying transporatbility methods | Estimated target parameter of the target population was statistical significantly different from the trial |
|---|---|---|---|---|---|---|---|---|---|---|
| Mollan et al., 2021(11) | Hazard Ratio | Yes | 2 (Efavirenz-containing vs. Efavirenz-free) | Yes | IOSW, multiple imputation, marginal structural Cox model | Not stated | No | The estimated HR for efavirenz affecting suicidal thoughts/behaviors was lower in the target population than in trials (transported: HR=1.8 vs. trials: HR=2.3). | Transporting trial outcomes to a target population was beneficial for examining external validity. The influence of efavirenz on suicidal thoughts/actions was observed but diminished in the target population. | Yes |
| O'Donnell et al., 2022(16) | Hazard Ratio | Yes | 3 (Two-unit UCB, Haplo-BM, Haplo-PB) | Yes | Multivariate analyses using Cox proportional hazards models | Not stated | No | There was no significant difference in 5-year PFS and OS between trial participants and non-participants, confirming the findings applicability. | The study confirmed the generalizability of the RCT results to real-world settings, indicating a preference for haploidentical transplants over cord blood with reduced intensity conditioning regimens. | Not stated |
| Nagasaka et al., 2022(13) | Hazard Ratio | Yes | 2 (including the control) | Yes | IPSW, propensity score modeling, entropy balancing | Not stated | No | After adjustments, the study indicated SPP maintained higher PFS than PPP, with other efficacy and safety results remaining consistent. | The study found statistically significant differences in PFS benefit. This study was a significant contribution to understanding how clinical trial results from one geographical location (China) could be translated and applied to another (USA) using advanced statistical methods. | Yes |
| Basu et al., 2023(10) | Mean difference | Yes | 3 (OAP-OAP, OAP-PP, PP-PP) | Yes | IPSW, double-lasso regression | Not stated | Yes | Adjusted treatment effects from the DREaM study for the real-world MMC cohort revealed that OAP-PP and PP-PP treatment strategies could notably decrease psychiatric hospitalizations in comparison to OAP. | The study showcased the practical implications of initiating PP therapy early, as seen by the findings of the DREaM trial. It indicated that using PP treatment strategies could effectively decrease mental hospitalizations among Medicaid recipients who have recently developed schizophrenia. | Yes |
| Wei et al., 2023(7) | Odds ratio | Yes | 2 interventions per trial (4 in total) | Yes | Regression-based transportation, IOSW | Not stated | No | Regression-based and IOSW methods can transport trial effect estimates to real-world settings with only moderate precision decreases. | The study demonstrates that HF trial results can be transported to real-world registry data, enhancing their applicability and representativeness without significant loss of precision. | No |
| Lee et al., 2021(15) | Mean difference | Yes | 2 (modular arm vs. monoblock arm) | Yes | IPSW, logistic regression, weighted linear regression. | Not stated | Not stated | Observed treatment effects in the trial were applicable across the broader UK population with intracapsular hip fractures. | The modern modular hemiarthroplasty did not provide additional benefit over the traditional monoblock in terms of quality of life and length of hospital stay when implemented nationwide in the UK. This study highlighted the importance and feasibility of | No |
| | | | | | | | | | assessing the generalizability of RCT findings to wider populations. | |
| Cook et al., 2023(17) | Hazard Ratio | Yes | 2 (Buprenorphine and Extended-Release Naltrexone) | Yes | IPSW, Cox proportional hazards model | Not stated | No | Stimulant use impeded the initiation of MOUD in both RCT and real-world populations, with effect size varying by population. | This study addressed interventions targeting stimulant use among patients with OUD was crucial for improving MOUD initiation rates, especially in real-world settings like rural areas with high rates of injection drug use. | Yes |
| Bengtson et al., 2016(12) | Mean difference | Yes | 2 (MBC intervention vs. enhanced usual care) | Yes | IPSW | Not stated | No | Standardization to the target population reduced the intervention effect by 1.2 points, indicating MBC may be less effective in routine care yet remains advantageous. | The study showed that MBC may have been less effective when implemented among HIV-infected depressed adults in routine care compared to RCT settings, but still reduced depression, which highlighted the need for more effective depression interventions for HIV-infected adults. | Yes |
| Doss et al., 2023(8) | Incidence | Yes | 2 (real-world PET and CAPSIL PET cohort) | Yes | Comparison of outcomes between real-world and CAPSIL cohorts, chi-square or Fisher's exact tests, Mann Whitney U test | Not stated | No | Similar feasibility and effectiveness in using PET for CMV disease prevention were noted in both real-world and RCT settings. | PET was feasible and effective for CMV disease prevention in high-risk D+R- LTxRs in real-world settings, with similar outcomes to those in the RCT. This study provided evidence supporting the generalizability and implementation of PET for CMV disease prevention in high-risk D+R- LTxRs across diverse settings. | No |
| Dahabreh et al., 2019(6) | Relative risk | Yes | 2 (surgery plus medical therapy vs. medical therapy alone) | No | Logistic regression models for trial participation, treatment assignment, and outcome probability; IPSW; augmented IP weighting | Not stated | Not stated | Similar 10-year mortality risks for surgery and medical therapy were observed between the target population and randomized individuals, indicating the generalizability of trial results. | The study supported the feasibility of generalizing causal inferences from RCT individuals to all trial-eligible individuals using baseline covariate data, treatment, and outcome information from randomized participants, with implications for enhancing the relevance of trial findings to broader populations. | No |
| Dahabreh et al., 2020(3) | Relative risk | Yes | 2 (surgery plus medical therapy vs. medical therapy alone) | No | Logistic regression for trial participation, treatment assignment, and outcome probability; IPSW; doubly robust approaches combining outcome and trial participation modeling | Not stated | Not stated | Estimated 10-year mortality risks for surgery plus medical therapy and medical therapy alone were similar among nonparticipants and trial participants, suggesting trial results are generalizable to a broader population of eligible nonparticipants. | The methods reviewed facilitated the extension of causal inferences from RCTs to target populations, using CASS as an illustrative example. The study revealed that the RCT results comparing coronary artery surgery plus medical therapy to medical therapy alone can be applied to a wider population of patients with chronic coronary artery disease. | No |
| Robertson et al., 2024(18) | Relative risk | Yes | 2 (surgery plus medical therapy vs. medical therapy alone) | Yes | G-formula, weighting, and augmented weighting estimators | Not stated | Yes | Subgroup-specific effects highlighted potential advantages of combining surgery with medical therapy over medical therapy alone, especially in patients with | Subgroup analyses using generalizability and transportability methods identified potential benefits of treatments in specific patient subgroups, highlighting the importance of considering | Not stated |
| | | | | | | | | prior myocardial infarctions, with estimates varying by estimation method. | heterogeneity of treatment effects in clinical and policy decision-making. | |
| Cole et al., 2010(9) | Hazard Ratio | Yes | 2 (including the control) | Yes | IPSW, Cox proportional hazards model | Not stated | No | Trial inferences were relevant to the target population, though diminished by 12%, based on accurate modeling of selection determinants. | The study illustrated a model-based method to standardize trial results to a specified target population. | Yes |
| Happich et al., 2020(14) | Hazard Ratio | Yes | 2 (including the control) | Yes | IPSW | Not stated | No | Inverse propensity score reweighting analysis indicated a marginally smaller effect of pemetrexed versus gemcitabine, with increased uncertainty and no statistical significance. | Inverse propensity score reweighting offered a novel approach to estimating expected treatment benefit in a broader real-world target population, addressing generalizability concerns at the time of regulatory approval/marketing authorization. | No |
| Butala et al., 2022(5) | Rate Difference | Yes | 2 (extended vs. standard DAPT duration) | Yes | IOSW | Not stated | No | The extended use of DAPT did not show clear benefits in reducing stent thrombosis, myocardial infarction, and MACCE in the contemporary population, while the risk of increased bleeding remained. | The differences between the populations tested and those observed in current clinical practice indicated that the advantages of prolonged DAPT may have restricted applicability, while the risks were more severe in the modern settings. | No |

*CMV, cytomegalovirus; D+R-, donor positive/recipient negative; DAPT, dual antiplatelet therapy; DREaM, Disease Recovery Evaluation and Modification study; HR, hazard ratio; Haplo-BM, HLA-haploidentical bone marrow; Haplo-PB, HLA-haploidentical peripheral blood; IOSW, inverse odds of sampling/scoring weights; IPSW, inverse propensity sampling/score weighting; LTxRs, lung transplant recipients; MACCE, major adverse cardiac and cerebrovascular events; MMC, medicaid managed care; MOUD, medication for opioid use disorder; OAP, oral antipsychotic therapy; OS, overall survival; PB, peripheral blood; PFS, progression-free survival; PP, paliperidone palmitate; PPP, placebo + pemetrexed + platinum; RCT, randomized controlled trial; SPP, sintilimab + pemetrexed + platinum; UCB, umbilical cord blood.